\documentclass[12pt,preprint]{aastex}
\usepackage{amsmath,amssymb}
\begin{document}
\title{Starburst Galaxies: Why the Calzetti Dust Extinction Law?}

\author{J\"org Fischera, Michael A. Dopita \& Ralph S. Sutherland}
\affil{Research School of Astronomy \& Astrophysics,
Institute of Advanced Studies, The Australian National University,
Cotter Road, Weston Creek, ACT 2611
Australia}
\email{fischera,mad,ralph@mso.anu.edu.au}

%\doublespace

\begin{abstract}
The empirical reddening function for starburst galaxies generated by Calzetti
and her co-workers has proven very successful, and is now used widely in the
observational literature. Despite its success, however, the physical basis
for this extinction law, or more correctly, attenuation law remains weak.
Here we provide a physical explanation for the Calzetti Law based on a
turbulent interstellar medium. In essence, this provides a log-normal
distribution of column densities, giving a wide
range of column densities in the dusty foreground screen.
Therefore, extended sources such as starburst regions or HII regions seen
through it suffer a point-to-point stochastic extinction and reddening.
Regions of high column densities are ``black" in the UV, but translucent in the 
IR, which leads to a flatter extinction law, and a larger value of the total to selective
extinction, ${\rm R_V}$.
We fit the Calzetti Law, and infer that the variance $ \sigma $ of the
log-normal distribution lies in the range $0.6\le \sigma \le 2.2$. The
absolute to selective extinction ${\rm R_V}$ is found to be in the range
$4.3$ to $5.2$ consistent with ${\rm R_V}=4.05\pm0.80$ of the Calzetti Law.
\end{abstract}

%\keywords{galaxies: starburst --- ISM: general --- ISM: dust}

%\newpage

\section{\label{Intro}Introduction}
As \citet{Calzetti2001} has made clear, the {\it extinction} curve of
a single star by foreground dust should be distinguished from the
{\it attenuation} curve of an extended object such as starburst galaxy
or an \ion{H}{2} region seen through a foreground dusty medium. The
extinction for a single star depends simply on the column density and
is determined by both the absorption and the scattering properties of the
grains in the foreground dusty screen, since $\kappa_{\rm ext} =
\kappa_{\rm abs} + \kappa_{\rm ext}$.
In the case of an extended object the emitted light suffers an effective
attenuation which depends strongly
%on the dust distribution in the foreground screen
on the relative distribution of dust and emitting stars.
%and the distribution of the emitters in respect to this screen.
In addition some of the measured flux can originate from light scattered
into the observed direction. For all these reasons, the effective attenuation
will tend to be less than extinction we might expect from a simple foreground
screen,
and the intrinsic wavelength dependence of the attenuation might be expected
to differ from the extinction curve of a single star.

 From an observational viewpoint, an accurate knowledge of the appropriate
attenuation curve for star-forming galaxies fundamental in helping us
to understand the formation and evolution of galaxies. Very extensive use
has been made of the Madau Plot \citep{Madau1995}, in which the estimated
star formation rate per unit co-moving volume of the Universe is plotted
against red-shift. For high redshift galaxies, one of the principal means
of populating this diagram is through a measure of the UV continuum, which
should scale as the star formation rate (SFR) in galaxies with high specific
star formation rates. However, these measures depend critically upon the
dust attenuation corrections. These are very uncertain; there are claims
that a typical $z=3$ galaxy suffers a factor of ten extinction at 
1500\AA\ in the
rest--frame of the galaxy \citep{Meurer1999,Sawicki1998}. Others suggest more
modest corrections \citep[e.g.][]{Trager1997}. In normal disk galaxies,
\citet{Bell2001} find that although mean extinctions are modest, about
1.4~mag at 1550\AA, the scatter from galaxy to galaxy is large.
For nearby starburst galaxies attenuation and SFR are strongly correlated
\citet{Buat1999}, and such a correlation extends to high-redshift galaxies
\citep{Adelberger2000}. It is possible that many fainter galaxies at high
redshift are completely missed, resulting in large uncertainties in the derived
star formation rates.

%The measurement of the effective attenuation in the UV depends
%upon the measurement of the slope, $\beta$, of the UV continuum in the range
%1200-2600\AA~(${\rm f}_{\lambda} \propto {\lambda}^{\beta}$)
%\cite{Calzetti1994}. The intrinsic slope of the UV continuum depends
%on the star formation history, and can be determined from stellar spectral
%synthesis codes such as STARBURST99 \citep{Leitherer1999}, see, for example,
%\citet{Leitherer2002}. Roughly, $\beta \sim -2$ for continuous star
%formation \citep{Calzetti1994} and $\beta \sim -2.7$ for an recent
%instantaneous burst of star formation \citep{Meurer1995}.

%Using this approach,
Calzetti and her co-workers have been able to derive
empirical attenuation curves across the full near-IR to far-UV spectral range.
These have been given as local (piecewise) power-law fits of the form
\citep{Calzetti1997, Calzetti2000, Calzetti2001, Leitherer2002}:
\begin{equation}
k(\lambda) ={\rm A}(\lambda)/{\rm E(B-V)_*}=a +b/\lambda 
+c/{\lambda}^2 +d/{\lambda}^3,
\end{equation}
where $a$, $b$, $c$, and $d$ are constants in a given wavelength range and
${\rm A}(\lambda)$ is the attenuation in magnitudes at wavelength
$\lambda$, ${\rm E(B-V)_*}$ the color excess of the stellar continuum light.
The total to selective attenuation of the stellar continuum,
$R_{\rm V}={\rm A}(\rm V)/{\rm E(B-V)_*} =4.05 \pm 0.8$.
Experimentally, \citet{Calzetti1994} showed that the HII regions are more
heavily attenuated than the stellar continuum, presumably because these
are associated with more dusty regions of ongoing star formation.
Although, from galaxy
to galaxy, there are considerable variations from the Calzetti Law it has
nonetheless proved to be very useful and is widely used by observers.

The purpose of this paper is to remove some of the empirical
aspects of this fit, by deriving an analytic fit to the attenuation based
on a turbulent model of the interstellar medium, and using the interstellar
grain model of \citet{Weingartner2001}. We show that, within a certain range
of parameters, and neglecting environmentally-dependent dust properties,
the theoretical attenuation of a turbulent dust screen agrees well with
the Calzetti law. We also provide an estimate of how this attenuation law
might change as we observe starburst galaxies with a larger global
attenuation, such as are found in the high-redshift universe at the epoch
of galaxy formation.

\section{\label{MHD}The Extinction of a Turbulent Foreground Screen}
\subsection{Density Fluctuations}
The interstellar medium (ISM) cannot be considered as homogeneous in any
galaxy.
The phase of the ISM which influences the attenuation law is not
primarily the gravity-dominated molecular clouds or Bok Globules, which
are highly optically thick, but rather the more diffuse turbulent phase.
\citet{Calzetti1996} and \citet{Calzetti1997} have investigated the effect
that a two-phase clumpy medium would have on the attenuation law. This
model has clumps of identical optical thickness providing a certain covering
factor in the foreground dusty screen.
The result of this is generate a stochastic reddening which results in a
flatter attenuation law than the standard interstellar extinction law.
%The reason for this is that the screen is ``leaky" so that more
%radiation is transmitted by the screen in the UV than would be the case for
%a uniform screen.

This simple model shows how inhomogeneities in the ISM can provide a
qualitative explanation of why the Calzetti Law works. However, it is
very simplistic. We now understand that the inhomogeneities in the
non-gravity dominated phases of the ISM are likely to be the result of
turbulence. % which provides a set of clouds possessing a very wide and
%continuous range of local densities and column densities.
%It is the properties of an absorbing screen made up of such clouds 
%that we must
%seek to understand.
%All turbulence is due to energy injection on the large scale,
%which cascades down to small scale eddies within one eddy
%turnover time.
%In the case that the gas is inertial, the well-known
%Kolmogorov spectrum applies, giving a one-dimensional energy spectrum
%$E(k) \propto k^{-5/3}$.
In the diffuse phases, MHD turbulence
may apply, with the magnetic field being an important contributor
to the total pressure.
%Indeed, using MHD turbulent models,
%\citet{ChoVish2000} have shown that field line stretching can amplify the
%magnetic fields up to the level of energy and pressure equipartition.
%In the weak field case the turbulent local magnetic field may be much
%larger than the external, mean magnetic field.
%Such MHD turbulence
%naturally leads to strong density fluctuations.
%Regions of high density
%and gas pressure, but small magnetic field pressure are matched in total
%pressure by other regions of low density and gas pressure, but with high
%magnetic pressure.
MHD turbulence can provide a natural explanation
for the famous FIR:radio correlation in galaxies \citep{Groves2002}.

All turbulence is characterised by a wide range of densities. As shown by
simulations of compressible hydrodynamic turbulence the density
distribution is well described by a log-normal distribution if the turbulence
is approximately isothermal (see \citet{Ostriker2001} and references therein)
where the log-normal distribution is given by:
\begin{eqnarray}
%\alignleft
& & p(\ln(\rho))= \frac{{\rm d}p(\ln(\rho))}{{\rm d}\ln(\rho)}
=\frac{1}{\sqrt{2\pi}\sigma}e^{{-x^2}/{2\sigma^2}} \nonumber \\
& {\rm with}& x =\ln(\rho)-\ln(\rho_0).
\end{eqnarray}
Here, the mean $\left<\rho\right>
=\int{\rm d}(\ln(\rho))\, \rho\,p(\ln(\rho)) = \rho_0\,e^{\sigma^2/2}$.
\citet {Nordlund1999} have shown analytically that the density distribution
in supersonic and isothermal turbulence is a log-normal distribution.
In principle the distribution becomes wider with increasing Mach number ${M}$.
In particular, for unmagnetised forced turbulence \citet{Nordlund1999} and
\citet{Padoan1997} found that the variance is correlated to the Mach number by:
\begin{equation}
   \label{sigmacorrelation}
   \sigma^2 = \ln\left(1+\beta^2\,M^2\right)
\end{equation}
with $\beta\approx0.5$. However, in general, no simple dependence between
Mach number and density contrast has been found \citep{Ostriker2001}.

As discussed by \citet{Ostriker2001}
the distribution of column densities $N=\int {\rm d}l\,\rho(\vec r)$ for an
isothermal turbulent medium is also approximated by a log-normal distribution.
Thus, in a turbulent
medium, as long as the dust size distribution and composition (which
determines the absorption coefficient, $\kappa_{\lambda}$, at wavelength
$\lambda$) is not a strong function of density, then a log-normal
distribution should also characterise the effective optical depth,
$\tau_{\lambda}=\int {\rm d}l\,\rho(\vec r)\,\kappa_{\lambda}
=\kappa_{\lambda}N$.
When the correlation length of the turbulence is much less than the scale
thickness of the absorbing layer, the contrast in column density will 
be smaller
than the density contrast \citep{Ostriker2001}. Thus, if equation
(\ref{sigmacorrelation}) is valid, the contrast in column density can 
be used to
obtain a \emph{lower} limit of the Mach number.

\subsection{Dust and Attenuation Model}
We have adopted the \citet{Weingartner2001} dust model with Galactic
abundances and $R_{\rm V}=3.1$ as providing the (currently) most
sophisticated theoretical fit to the local extinction curve.
In the more extreme UV radiation fields encountered in starburst galaxies,
this grain model may not be an appropriate choice, because the
small carbonaceous grains containing large amounts of PAHs
are likely to be destroyed. The main effect of this will be to suppress
the  2200\AA\ feature, which in starburst galaxies is
%observed to be 
either very weak or absent.

To account for extinction it is important to know the relative distribution
of stars and gas.
%how dust and stars are distributed inside the galaxy.
Since the foreground dust screen model has proved to be most successful in
describing the attenuation properties of galaxies \citep{Calzetti2001},
here we assume that the attenuation is caused by a turbulent
screen in front of the emitting stars. Other geometries will be
investigated in a later paper.
Because of the distribution of column densities in this screen the star light
suffers, either by absorption or either by scattering at dust
grains, a range of attenuation $e^{-\tau}$.
The ratio of the observed to the emitted light of all stars can be 
attributed to
an effective
optical depth $\tau_{\rm eff}$. Its value is given by
the averaged attenuation:
\begin{eqnarray}
    \tau_{\rm eff}
&= -\ln{\left(\int{\rm 
d}\ln(\xi)\,p(\ln(\xi))\,e^{-\xi\left<\tau\right>}\right)}
\nonumber\\
&= -\ln\left(\int{\rm d}y\,p(y)\,e^{-e^y\left<\tau\right>}\right),
\end{eqnarray}
where $p(x)$ is the log-normal distribution of $\tau/\left<\tau\right>=\xi=e^y$
and $\left<\tau\right>$ the averaged optical thickness of the screen.
%It will be shown how the width $\sigma_{\tau/\left<\tau\right>}$ of
%the distribution in $\tau$ and
%the mean optical thickness $\left<\tau\right>$ determine the observed quantity
%$\tau_{\rm eff}$. Due to the exponential behavior of the attenuation
%the deviation
%of the effective optical thickness to the averaged optical thickness
%increases with
%increasing extinction coefficient $\kappa_{\lambda}$ of the dust.
%By taking the extinction curve determined for galaxies where an
%approximation of a foreground dust screen for the dust extinction is 
%appropriate
%the width of the column density distribution can be obtained which%
%may in turn provide some observational constraints on the nature of 
%the turbulence
%in the ISM of these galaxies.

\section{A Fit to the Calzetti curve}

\begin{figure}
\scalebox{1.}[1.]{
\plotone{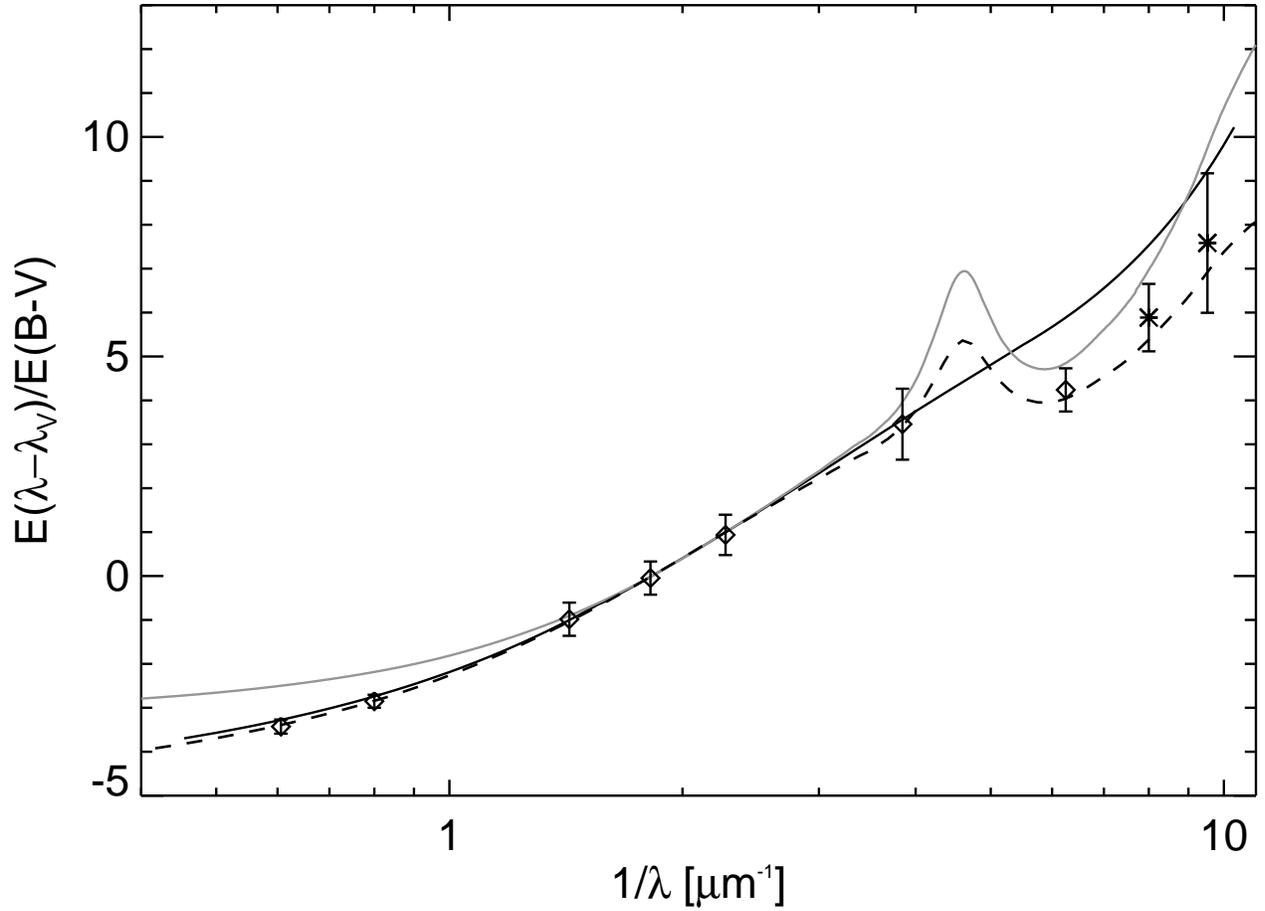}}
		\caption{\label{fig1} The theoretical fit to the Calzetti attenuation
   curve. The Calzetti curve itself is the dark solid line, and the data
   points used in the fit from Table(1) are shown along with their errors.
   The grey curve is the mean Milky Way extinction curve, and
   our least-squares fit to a turbulent foreground dust screen
   is shown as a dashed line.}
\end{figure}

The analytical form of the Calzetti attenuation curve given by equation (1)
is itself a fit derived from a large number of observations
made in different wavebands. To model this extinction curve, and to obtain
some idea of the uncertainties in both the data and our fitted model, we need
to fit to data points from which the Cazetti curve is derived. This proved to
be rather difficult, because the errors are not very well constrained in much
of the fitting procedures used by Calzetti. We have adopted the following
compromise. In the range of $0.25$ to $1.65~\mu{\rm m}$ we have used two data
sets given in \citet{Calzetti1997}. These give the selective obscuration at
seven wavelengths, $E(\lambda-2.2\,\mu{\rm m})_*/E(B-V)_{\rm HII}$ derived from
the ${\rm E(B-V)}_{{\rm H}_{\alpha}/{\rm H}_{\beta}}$ and the
${\rm E(B-V)}_{{\rm H}_{\beta}/{\rm Br}_{\gamma}}$ correlations.
First we combined both data sets ($f_1, f_2$) with their respective
uncertainties $\sigma_1$ and $\sigma_2$ by taking an weighted average:
$\left<f\right> =
\left(f_1/\sigma_1^2+f_2/\sigma_2^2\right)/(1/\sigma_1^2+1/\sigma_2^2)$.
After applying their recommended correction to the stellar selective
extinction,
${\rm E(B-V)_*=0.44\,E(B-V)_{\rm HII}}$, we found that almost all
these data points
lie well above the values of the selective obscuration obtained from
the analytic
formula (1). To allow a comparison of the model with the Calzetti curve we
scaled the data by a factor 0.785 which minimized the $\chi^2$ of the
data values in the range from 0.26 to $1.65~\mu{\rm m}$ with respect to the
analytical curve.
The derived values are summarised in table \ref{measurements}. With
the exception
of the measurement at $0.16~\mu{\rm m}$ these values are in excellent agreement
with the analytical curve of   fig (\ref{fig1}).

%\clearpage

\begin{table}
    \caption{The data points used to fit the extinction curve of
star-burst galaxies.
Here, $k'(\lambda)={\rm A'}_{\lambda}/{\rm E(B-V)_*}$ and
$\Delta k'(\lambda)=\Delta\,{\rm A'}_{\lambda}/{\rm E(B-V)_*}$.}
\label{measurements}
    \begin{tabular}{l|cc|ccccccc}
		\hline
		\hline
    $\lambda$ [$\mu$m]& 0.105\tablenotemark{1} & 0.125 & 0.16\tablenotemark{2} &
0.26 & 0.44 & 0.55 & 0.70 & 1.25 & 1.65 \\
    \hline
    $k'(\lambda)$ & 11.71 & 10.02 & 8.37 & 7.59 & 5.07 & 4.08 & 3.15 &
1.28 & 0.70 \\
    $\Delta k'(\lambda)$ & 1.59 & 0.77 &0.49 & 0.81& 0.46& 0.38& 0.38 &
0.15&0.16\\
		\hline
		\hline
    \end{tabular}
\tablenotetext{1}{Based on \citet{Leitherer2002}}
\tablenotetext{2}{Based on \citet{Calzetti1997}}
\end{table}

%\clearpage

To extend the fitting region to wavelengths shorter than $0.16~\mu{\rm m}$ we
have added two further points at $0.125$ and $0.105~\mu{\rm m}$ based
on the HUT measurements of the selective obscuration
${{\rm E}(\lambda-0.15~\mu{\rm m})/{\rm E(B-V)}_*}$ given by
\citet{Leitherer2002}.
The absolute values were found by normalising to the point at
$0.16~\mu{\rm m}$.
As a consequence of this normalisation, these points lie below the
$k(\lambda)$ of the Calzetti curve.
This difference illustrates how sensitive the curve is to errors
in the piecewise fitting procedure. We will refer to the $k(\lambda)$ values
given in table (\ref{measurements}) as $k'(\lambda_i)$.

\begin{figure}
\scalebox{1.}[1.]{
\plotone{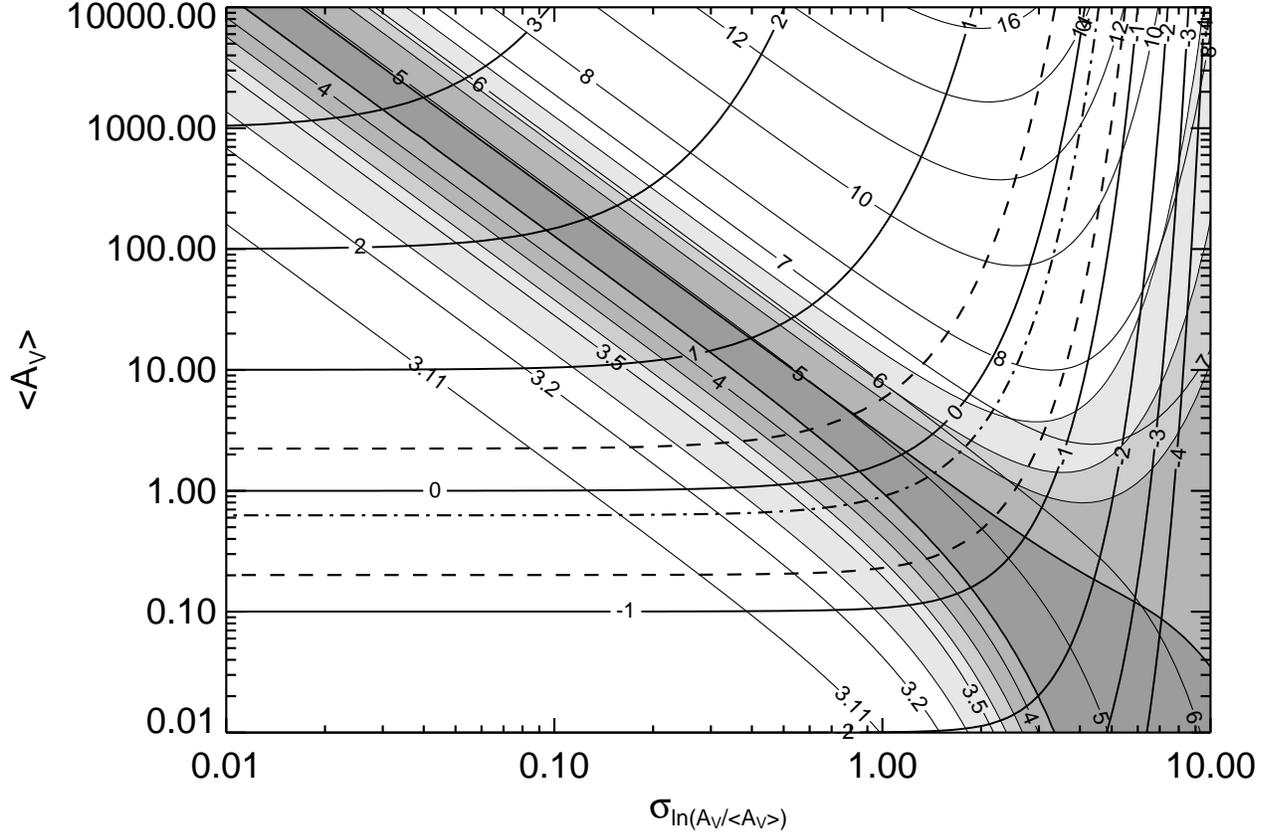}}
\caption{\label{fig2} The observable quantities ${\rm A_V}$ and
${\rm R_V}$ are here plotted as functions of the theoretical
parameters of the foreground screen, ${\rm \left<\rm A_V \right>}$ and
the variance $\sigma$ of the log-normal distribution of the optical 
depths $\tau/\left<\tau\right>={\rm A_V/\left<A_V\right>}$.
The shaded contours represent the fitting to the Calzetti curve.
The contours are the 68, 90, 95 and 99\% confidence levels.
The dashed, dashed dotted and dashed line are the lower, mean, and 
upper values of the
attenuation observed in the galaxies.
These are used to define the appropriate $\sigma$ values
which characterize the variation of column densities of the dust screen.}

\end{figure}

In our fitting procedure, we used the theoretical turbulent cloud structure
discussed in the previous section, described by a mean extinction
$\left<{\rm A}_{\rm V}\right>=2.5\left<\tau_{\rm V}\right>\log(e)$ and variance
$\sigma_{\ln(\tau/\left<\tau\right>)}$ of the log-normal distribution
of the optical
thickness $\tau$. These theoretical parameters together determine a unique
attenuation curve which we derived using the dust model of
\citet{Weingartner2001}.
In practice, we find that all attenuation curves having the same
effective value of
the total to selective extinction, ${\rm R}_{\rm V}$, are almost identical, and
that each value of ${\rm R}_{\rm V}$ can be represented as a curve on the
$\left <{\rm A}_{\rm V}\right> : \sigma_{\ln(A_{\rm V}/\left <{\rm
A}_{\rm V}\right>}$
plane. Likewise, lines of constant effective attenuation of the screen,
${\rm A}_{\rm V}$, can also be represented by curves on this plane {\emph see}
fig (\ref{fig2})). Note that for small values of $\sigma$, the effective
attenuation is the same as the mean extinction, as it should be for an (almost)
uniform medium.

To determine the parameters required to fit the Calzetti curve, we fitted the
obscuration $k'(\lambda_i)$ using a $\chi^2$-fit.
To avoid the uncertainties in the absolute value in the analytical formula
which had been measured independently in respect to the selective obscuration
\citep{Calzetti2001}, we retain an additive constant as an additional
free parameter.

The result of the $\chi^2$-fit is shown as
contours in figure (\ref{fig2}). The best fit has reduced $\chi^2$ of $0.048$. 
This is an excellent fit and suggests the data errors may have been 
overestimated. 
The corresponding attenuation curve is shown in figure (\ref{fig1}).
Below a variance of $\sigma_{\ln(\tau/\left<\tau\right>)}\approx 2$ the range
of solutions inside a certain confidence level is consistent with a restricted
range of ${\rm R_V}$. The solutions of $68\%$ confidence constrain  ${\rm R_V}$
in the range $4.3 \le {\rm R_V} \le 5.2$. 
This is consistent with the value given
for the analytical extinction curve (${\rm R_V}=4.05\pm0.8$).

The range of physical solutions are further constrained by the absolute
attenuation characterising the observed galaxies; $0.25 < {\rm A_B} < 2.78$
with a mean of $0.78$, or $0.20 < {\rm A_V} < 2.23$ \citep{Calzetti2001}.
  From figure (\ref{fig2}), these values, combined with the limits derived
above on ${\rm R_V}$ restrict $\sigma$ to lie in the range $0.6 - 
2.2$ with a most probable value $\sigma \sim 1.0$.

Through the IR to near-UV, the quality of our theoretical fit to the Calzetti
curve is remarkable. In the UV, the obvious deficiency is the inclusion of the
2200\AA~absorption feature, which reflects the deficiencies in our simple
grain model, which does not take account of
%although predicted weaker than the Milky Way
%mean extinction law, is still present, and much stronger than seen in real
%star-burst galaxies. As mentioned, this is likely explained by
the destruction
of small carbonaceous grains in starburst environments.
%Even though the values of
%the theoretical curve deviate in the UV from the values of the analytical
%formula the curvature of both theoretical and observed curves is 
%almost the same.

\section{Discussion \& Conclusions}

We have demonstrated that the theoretical attenuation of light through a
turbulent dusty foreground medium can provide an excellent fit to the empirical
Calzetti attenuation curve, provided that $0.6 < \sigma_{\ln(A_{\rm V}/
\left <{\rm A}_{\rm V}\right>} < 2.2$ and  $4.3 \le {\rm R_V} \le 5.2$.
If formula (\ref{sigmacorrelation}) is correct, the inferred
contrast in column densities implies a \emph{minimum} value of the Mach number
in the dusty layers in the range of $1.3$ to $22$.

These numbers can be compared with estimates appropriate to the warm neutral
medium (WNM) and the cold neutral medium (CNM) of galaxies. When the vertical
velocity  dispersion of quiescent disk galaxies is measured it is found to
be remarkably  constant.  The \ion{H}{1} observations, which are sensitive
to the WNM, give vertical velocity dispersions between 7 and 10 km~s$^{-1}$
\citep{Kruit1982, Shostak1984,Kim1999,Sellwood1999}. The 3D velocity
dispersion will be $\sqrt3$ times these values. Assuming that the WNM
has a temperature of order 6000K, this implies $M \sim 1.8$ in this
component. For the warm ionised medium (WIM), the estimated Mach number is
very similar. For the CNM, we can use estimates based on observations of
CO, which give vertical velocity dispersions of $6-8$ km~s$^{-1}$
\citep{Combes1997}. Assuming a temperature $\sim100$K in this component, we
have $M \sim 12$. Starburst galaxies are characterised by somewhat higher
velocity dispersions so we would expect that for these the Mach numbers are
somewhat higher. Nonetheless, the agreement of these estimates with the
numbers provided by the dust attenuation law fit is very pleasing, and
gives greater confidence that the ISM density structure in the dusty ISM is
determined by turbulent processes acting in the relevant phases.

It is somehow surprising that the attenuation model works so well even
without the inclusion of scattered light. This is in contradiction
to expectation that the losses due to scattering cancel - every photon
scattered out of the viewing direction being compensated by another
photon scattered into this direction. However, it is easy to understand
that the out-scattered photons cannot be totally compensated by the
in-scattered photons because the distance the scattered photons travel
through the dust screen is different and the total optical depth experienced
by the photons which are scattered into the viewing direction is
always larger.

A further reason why the scattering does not contribute significantly to
the attenuation may be caused by the clumpiness of the screen itself. It is
true that the absolute contribution of the scattered light to the total
transmitted light increases if the medium becomes clumpy. However, in
comparison with the directly transmitted light this is a negligible
contribution. Due to the log-normal distribution, the optical depth
is mostly much smaller than the average optical thickness. It is in the high-density
clouds where most of the extinction and scattering extinction occurs.
The column density in such clouds is so high that photons, if they are not
absorbed immediately, will suffer multiple scattering events until they are
finally absorbed. Only photons scattered at the outskirts of these dense
regions may escape and a small fraction will be scattered into the viewing
direction. This will result in a slightly smaller extinction than that
derived here.

%To the extent that the turbulent structure of the ISM is invariant from
%galaxy to galaxy, these models predict that we would expect to find a
%correlation between ${\rm R_V}$ and ${\rm A_V}$. This is driven by 
%the $\sigma$
%of the ISM.
To the extent that the turbulent ISM structure is invariant
from one starburst galaxy to another, these models predict a correlation
between ${\rm R_V}$ and ${\rm A_V}$ provided that the contrast in the column
density is the same.
%However, the $\sigma$ in the column density variations will not
%be the same as the $\sigma$ in the local density.%
As we go to a higher mean column density, the contrast in 
column density will decrease, as long as the local density contrast and the 
mean density remains constant.
In a forthcoming paper, we show that a higher column
density leads to a further flattening of the extinction curve and to
an increase of ${\rm R_V}$, mild in comparison to the
increase in $\left<{\rm A_V}\right>$. This flattening in the reddening curve
would be much stronger in more dusty galaxies or in violent systems with high
Mach numbers, such as merging or collapsing galaxies. Thus we would expect
that high redshift galaxies with strong star formation will have both large
${\rm R_V}$ and ${\rm A_V}$. This remains to be confirmed by observation.

\begin{acknowledgements}
The authors like to thank Bruce Elmegreen as a referee for his constructive
comments. M. Dopita acknowledges the support of the Australian 
National University
and of the Australian Research Council through his ARC Australian Federation
Fellowship. All authors acknowledge financial support for this 
research through ARC
Discovery project DP0208445.
\end{acknowledgements}

\clearpage

\clearpage

\clearpage

\clearpage

\end{document}